\documentstyle[12pt]{article}
\newcommand{\bfvec}[1] {\mbox{} {\mbox{}} {\mbox{\boldmath$ #1$}}}

\renewcommand{\theequation}{\arabic{section}.\arabic{equation}}
\newcommand{\ds}{\displaystyle}
\title{The Interference Rate of Radiation of Two Charges in Circular Motion}
\author{D. Villarroel and R. Rivera\\
\small\rm  Departamento de F\'\i sica \\
\small\rm Universidad T\'ecnica Federico Santa Mar\'\i a\\
\small\rm Casilla 110-V. Valpara\'\i so--Chile
}

\date{}

\begin{document}
\maketitle

\abstract{We present an exact formula for the computation of the interference
rate of radiation in the case of two charges revolving with constant angular
velocity at opposite ends of a diameter in a fixed circle.  The formula is
valid for arbitrary velocities of the charges, and can be easily studied by
numerical methods, even for velocities very close to the velocity of light.
For ultrarelativistic motion, the interference rate of radiation behaves as
$\ln(1 - v^2/c^2)^{-1/2}$, which contrasts with the behavior
$(1-v^2/c^2)^{-2}$ for the rate of radiation for one charge in circular
motion.  This is the first exact calculation for the interference rate of
radiation of two relativistic charges, and it is useful in connection with the
old controversy about the correctness of the Lorentz-Dirac equations of motion
for more than one charge.

}

\newpage
\setlength{\baselineskip}{1.5\baselineskip}{

\section{Introduction}

In the case of a one charge in arbitrary motion, the total rate of radiation
emitted at time $t$ is given by the Larmor's formula, which reads
\begin{equation}\label{e1}
\frac{d W_{{\rm rad}}}{dt} = \frac{2}{3} \, \frac{e^2}{c} \gamma^4
\left\{({\dot{\bfvec{\beta}}})^2 + \gamma^2 (\bfvec{\beta} \cdot
\dot{\bfvec{\beta}})^2 \right\},
\end{equation}
where ${\bf v}$ is the charge velocity, $c$ is the velocity of light, $\bfvec
{\beta} = {\bf v}/c$, \, $\dot{\bfvec{\beta}} = d\bfvec{\beta}/dt$, and
$\gamma = (1 - \beta^2)^{-1/2}$.  The simplicity of this formula, where the
variables $\bfvec{\beta}$ and $\dot{\bfvec{\beta}}$ are evaluated at the same
time as the total rate of radiation, reflects some special properties of the
point charge field.  There are basically two derivations of the Larmor's
formula (\ref{e1}).  In one of them \cite{1,2}, the total rate of radiation is
first calculated in the Lorentz rest frame of the charge, and equation
(\ref{e1}) is obtained using covariance arguments under Lorentz
transformation.  The other derivation is carried out directly in Minkowski
space, using the fact that the radiation of a point charge can be
characterized locally \cite{3,4,5}.  In particular, because of this property,
it is not necessary to go very far from the charge in order to calculate the
total rate or radiation.

In constrast with the one charge case, a formula for the total rate of
radiation of two charges under arbitrary motion is not known yet.
Furthermore, does not exist in the literature an exact formula for the total
rate of radiation for any special type of motion of two charges.  The source
of the technical difficulty comes, of course, from the fact that the fields of
the charges are explicit functions of the retarded times instead of the
laboratory time.  This point can also be illustrated by referring to the
already mentioned derivations of (\ref{e1}).  Thus, as emphazised by Landau \&
Lifschitz \cite{2}, for two charges there is generally no system of reference
in which both charges are at rest simultaneously.  And, on the other hand, the
local characterization of radiation does not work for more than one
charge\cite{6}.

Due to the superposition principle and the quadratic nature of the Poynting's
vector, the energy flux for two charges contains three terms.  Two of them are
associated with the field of each individual charge separately.  The third one
corresponds to an interference term that mixes the fields of both charges.  In
the evaluation of the total rate of radiation, the first two terms give rise
to a Larmor formula for each charge; so the real problem is the calculation of
the energy flux, across the surface of a sphere of very large radius, of the
interference term.

There have been only a few attemps to calculate the total rate of radiation
associated with the interference term.  Huschilt and Baylis
\cite{7} studied this radiation in the case of two identical charged
particles, that are moving in a straight line in head-on collision.  This
calculation is carried out with the help of the equation of motion for the
charges, and in addition it involves some kind of non-relativistic
approximations. Aguirregabiria and Bel \cite{8} studied the radiation of two
charges using a covariant formulation.  These authors calculated an integral
of the interference field over a circle, for a rather general motion of the
charges.  From this result, they elaborated a formalism, with the help of the
equations of motion and some additional assumptions, in order to obtain
succesive approximations for the total rate of radiation.  The interference
radiation has been studied also by Hojman et al \cite{9}, by means of a
covariant formalism.  However, these authors also introduce some kind of
non-relativistic approximations.

In this paper we present an exact formula for the total rate of radiation in
the case of two charges moving in a plane at the opposite ends of a diameter,
revolving at constant angular velocity in a fixed circular orbit. The main
motivation for this calculation is that it helps to solve an old controversy
about the correctness of the Lorentz-Dirac equations of motion for more than
one charge \cite{10}.  In fact, we have recently showed that, with appropriate
external fields, the Lorentz-Dirac equations describes the circular motion
under consideration in the case of two particles of equal charge and mass
\cite{11}.  Then, if we know the total
rate of radiation for this motion of the charges, we can check the consistency
of the Lorentz-Dirac equations with the energy conservation law.  We carried
out such an analysis in \cite{11}, and showed that this type of circular
motion allows us to see in a manifiest way the inconsistency of the
Lorentz-Dirac equations for more than one charge with the energy conservation
law.

The calculation is carried out directly in the laboratory frame.  This is more
appropiate than the covariant techniques, since the total rate of radiation
can be clearly visualized from a physical point of view, and the result is
free of any ambiguities whatsoever.  Our formula is an integral expression for
the total rate of radiation.  Unfortunately, the integral is too complicated
for closed analytical evaluation.  Nevertheless, it can be easily studied
using numerical methods, even for velocities near the velocity of light.
Furthermore, the integral can be evaluated approximately for ultrarelativistic
motion of the charges.  We find that the rate of interference radiation grows
as $(\beta^4/4) \; \ln\gamma\pi$ when $\beta$ tends to one.  This result
differs strongly with the behavior of the Larmor term of each charge, which
behaves as $\beta^4\gamma^4$ when $\beta$ is near one.  Thus, for example, if
we consider electrons of 500 Mev, the interference radiation is completely
negligible in comparison with the Larmor term, since this latter is of the
order of $10^{12}$, while the interference term is near $1$.  For low
velocities, however, it is known that the Larmor and the interference terms
are comparable \cite{12}.

In section II we evaluate the energy flux across the surface of a sphere
centered at the orbit center of the two charges, and show that this flux is
independent of the time at which it is evaluated, as well as of the radius of
the sphere.  In section III we present a power series expansion in $\beta$ of
the total rate of radiation up to $\beta^8$.  In section IV we derive an exact
formula for the interference radiation term.  In section V we present an
analytical approximation of the exact formula for the case of
ultra-relativistic motion.

\section{The interference rate of radiation}
\setcounter{equation}{0}

In the following we will be concerned only with the radiation of two charges
moving in a plane at opposite ends of a diameter, revolving at constant
angular velocity $\omega$, in a fixed circular orbit of radius $a$.  In this
case, because of the symmetries of the motion, it is possible to identify
without any ambiguities the total rate of radiation.  With this purpose in
mind we work directly in the laboratory frame, since this allows us to have a
clear physical picture of the radiation.

\begin{center}
\fbox{\hspace{3cm}FIGURE 1 \hspace{6cm}}
\end{center}

As it is shown in figure 1, our coordinate system is such that its origin
coincides with the center of the orbit, and the $X - Y$ plane is precisely the
orbit plane.  In this figure we have drawn the positions of the charges at an
arbitrary time $t$, and two spherical surfaces $\Sigma_1$, and $\Sigma_2$
centered at the origin of radii $r_1$ and $r_2$ respectively, with $r_2 > r_1
> a$, where $a$ is the orbit radius.  For a given time $t$, the electric and
magnetic fields ${\bf E}$ and ${\bf B}$ change in a very complicated way from
one point to another over the surface $\Sigma_1$; this is because the retarded
times of the two charges, which are in general different, change in a
complicated way with the position over $\Sigma_1$.  In particular then, the
Poynting vector ${\bf S} = (c/4\pi) {\bf E} \times {\bf B}$ is a complicated
function over $\Sigma_1$, and so it is the total flux across $\Sigma_1$ at
time $t$ that we are interested in.  Since the charges are moving jointly at
constant angular velocity, the position of them at different times looks the
same with respect to the whole surface $\Sigma_1$.  This means that the energy
flux across $\Sigma_1$, cannot depend on time.  In Section IV we present a
rigourous proof of this property, using the explicit form of the
electromagnetic field of the two charges.

Now, if we denote by $u({\bf x}, t)$ the energy density of the electromagnetic
field, and by ${\bf S}$ the Poynting vector $(c/4\pi) {\bf E} \times {\bf B}$;
then, since the domain $\Omega$ bounded by the two spherical surfaces
$\Sigma_1$ and $\Sigma_2$ of Fig. 1 is free of charges, the following
conservation law holds in it:
\begin{equation}\label{e2}
\nabla \cdot {\bf S} + \partial u/\partial t = 0.
\end{equation}
 From this equation we obtain
\begin{equation}\label{e3}
\int_{\Sigma_1} \left( {\bf S} \cdot \hat{\bf r} \right) d \Sigma_1 -
\int_{\Sigma_2} \left( {\bf S} \cdot \hat{\bf r} \right) d \Sigma_2 =
\frac{d}{dt} \int_\Omega u \left( {\bf x}, t\right) d^3x,
\end{equation}
where $\hat{\bf r} = (\sin \theta \cos \varphi, \, \sin\theta \sin
\varphi, \, \cos \theta)$ is the unit normal to  $\Sigma_1$ and $\Sigma_2$.
But, because of the symmetries of the motion of the two charges, it is clear
that the total energy contained in $\Omega$ is independent of time.  Therefore
the integral
\begin{equation}\label{e4}
\int_\Sigma \left( {\bf S }\cdot \hat{\bf r} \right) d\Sigma,
\end{equation}
over the surface of the sphere of radius $r$ is not only independent of time,
but it is also independent of the radius $r$.  In particular then, the
integral (2.3) represents the total rate of radiation that escapes to
infinity; and it can be evaluated over the surface of any sphere of arbitrary
radius $r$, with $r > a$.  Thus, for this special type of motion of the two
charges,a reminiscence of the local characterization of the radiation still
survives, in the sense that it is not necessary to go to infinity in order to
calculate the total rate of radiation.

We remark that the energy flux (2.3) is also independent of time and of the
radius of $\Sigma$, in the case of only one charge in circular motion with
constant velocity.  Therefore, if we denote by ${\bf E}_1$, ${\bf B}_1$, and
${\bf E}_2$, ${\bf B}_2$ the electric and magnetic fields of the charges $e_1$
and $e_2$ respectively, the integral
\begin{equation}\label{e5}
(cr^2/4\pi) \, \int_0^\pi \,\sin \theta d \theta \, \int_0^{2\pi}
\left({\bf{E}}_1 \times {\bfvec{B}}_2 + {\bf{E}}_2 \times {\bf {B}}_1
\right)
\cdot {\hat{\bf{ r}}} \,\textstyle{ d }\varphi,
\end{equation}
is independent of time and of the radius $r$.  The integral (2.4) represents
physically the rate of radiation due to the interference of the fields of both
charges, and it will be discussed in detail in section IV.

\section{Power series representation for low velocities}
\setcounter{equation}{0}

The total rate of radiation of several charges can be in principle calculated
by referring the retarded times of the charges, to the actual time $t$, by
means of a power series expansion in $1/c$.  The electric intensity of a group
of charges can be represented by the following formula when $r$ goes to
infinity \cite{12}.
\begin{equation}\label{e31}
\begin{array}{rcl}
{\bf E} & = & \ds{\frac{1}{rc^2}}
\left\{
\left[ \ds{\frac{d}{dt}} \left(\sum\limits_s e_s {\bf v}_s \right) +
\ds{\frac{1}{1!c}} \,
\ds{\frac{d^2}{dt^2}} \left( \sum\limits_s \, {\bf\hat{r}} \cdot
{\bf\bf{r}}_s e_s { \bf v}_s \right)\right.\right.\\ & & \\ & & + \cdots +
\ds{\frac{1}{(n-1)!}} \, \ds{\frac{1}{c^{n-1}}} \,
\left.\left. \ds{\frac{d^n}{dt^n}}
\left( \sum\limits_s ({\bf\hat{r}} \cdot {\bf r}_s)^{n-1} e_s
{\bf v}_s\right) + \cdots
\right] \times {\bf \hat{r}} \right\} \times {\bf\hat{r}},
\end{array}
\end{equation}
where ${\bf{\hat{r}}} = {\bf\hat{i}} \sin\theta \cos \varphi + {\bf\hat{j}}
\sin \theta \sin \varphi + {\bf\hat{k}} \cos \theta$, \, ${\bf\bf{r}}_s$
\, and \, ${\bf{v}}_s$ denote the position and velocity of the charge
$e_s$ at time $t$.  The far magnetic fields is ${\bf{B}} = {\bf{\hat{r}}}
\times {\bf{E}}$.

In our case we have
\begin{equation}\label{e32}
\begin{array}{rcl}
{\bf{r}}_1(t) & = & {\bf\hat{i}} a \cos \omega t + {\bf\hat{j}} a \sin
\omega t, \\
{\bf{r}}_2(t) & = & -{\bf\hat{i}} a cos \omega t - {\bf\hat{j}} a \sin
\omega t.
\end{array}
\end{equation}
For this motion we have computed the series (3.1) up to terms of power
$c^{-10}$, and the result is given in appendix $A$.  Using these formulae, the
calculation of the energy flux across the surface of a sphere of very large
radius $r$ is straightforward, but cumbersome.  The result is the following
\begin{equation}\label{e33}
\begin{array}{rcl}
\ds{\frac{dW_{\rm rad}}{dt}} & = & \ds{\frac{2}{3}} \,
\ds{\frac{e_1^2c}{a^2}} \beta^4 \left\{ 1 + 2 \beta^2 + 3\beta^4 + 4
\beta^6 + 5\beta^8 + \cdots \right\} \\
& & \\ & & + \ds{\frac{2}{3}} \, \ds{\frac{e_2^2c}{a^2}} \beta^4 \left\{ 1 + 2
\beta^2 + 3\beta^4 + 4\beta^6 + 5 \beta^8 + \cdots \right\} \\
& & \\ & & - \ds{\frac{4}{3}} \, \ds{\frac{e_1 e_2 c}{a^2}} \beta^4 \left\{ 1
-
\ds{\frac{14}{5}} \beta^2 + \ds{\frac{53}{7}} \beta^4 -
\ds{\frac{18556}{945}} \beta^6 +
\ds{\frac{515591}{10395}} \beta^8 + \cdots \right\}.
\end{array}
\end{equation}
The first two series correspond to power series expansions of the Larmor term
of each charge.  In fact, for a charge in circular motion with constant
velocity, the Larmor formula (1.1) is reduced to
\begin{equation}
\frac{2}{3} \, \frac{e^2 c}{a^2} \beta^4 ( 1 - \beta^2 )^{-2}.
\end{equation}

The series
\begin{equation}\label{e35}
-\frac{4}{3} \, \frac{e_1 e_2 c}{a^2} \beta^4 \left\{ 1 - \frac{14}{5}
\beta^2 + \frac{53}{7} \beta^4 - \frac{18556}{945} \beta^6 +
\frac{515591}{10395} \beta^8 + \cdots \right\},
\end{equation}
of (3.3) represents, of course, the rate of radiation associated with the
interference between the fields of both charges.  The first two terms of (3.5)
are already known \cite{12}.  As we will show, the series (3.5) gives the
total rate of interference radiation with an error less than 4\% for $\beta <
0.2$. Clearly, the error increases with $\beta$, and a power series in $\beta$
for the interference radiation is hopeless for $\beta$ near $1$, since a very
large number of terms would be needed in this region.

\section{An exact formula}

\setcounter{equation}{0}

We will consider now the energy flux of the interference term across the band
between the angles $\theta$ and $\theta + d\theta$ over the surface of the
sphere of radius $r$ at time $t$, that is, the contribution

\begin{equation}\label{e41}
\int\limits_0^{2\pi} \left( {\bf E}_1 \times {\bf B}_2 + {\bf E}_2 \times
{\bf B}_1\right) \cdot {\bf\hat{r}} d\varphi,
\end{equation}
to the rate given in Eq. (2.4).  The electric field ${\bf E}_1$ generated by
the charge $e_1$ is given by the well-known Lienard-Wiechert formula \cite{13}
\begin{equation}\label{e42}
{\bf E}_1 ({\bf x}, t) = e_1 \left[ \frac{ {(\bf\hat{n}}_1 -
{\bfvec{\beta}}_1)(1 - \beta_1^2) } {\kappa_1^3 R_1^2}
\right]_{{\rm ret}}
+ \frac{e_1}{c} \left[
\frac{
{\bf\hat{n}}_1}{\kappa_1^3 R_1} \times \left\{( {\bf{\hat{ n}}}_1 -
{\bfvec{\beta}}_1) \times {\dot{\bfvec{\beta}}}_1
\right\} \right]_{{\rm ret}}.
\end{equation}
The corresponding magnetic induction $B_1$ is
\begin{equation}\label{e43}
{\bf B}_1 = {\bf\hat{n}}_1 \times {\bf E}_1.
\end{equation}
In equation (4.2) and (4.3), ${\bf\hat{n}}_1$ is the unit vector that
points from the retarded position ${\bf r}_1(t_1)$ of charge $e_1$, associated
with the point ${\bf x}$ and time $t$, to the detection point ${\bf x}$.  The
vectors ${\bfvec{\beta}}_1$ and $\dot{\bfvec{\beta}}_1$ have already been
defined in connection with formula (1.1); but now they have to be evaluated at
the retarded position ${\bf r}_1(t_1)$ of the charge $e_1$.  Moreover, $R_1$
represents the distance between the detection point ${\bf x}$ and the retarded
position ${\bf r}_1 (t_1)$ of charge $e_1$; and $\kappa_1$ denotes the
following positive number
\begin{equation}
\kappa_1 = 1 - {\bf\hat{n}}_1 \cdot {\bfvec{\beta}}_1.
\end{equation}
The electric field ${\bf E}_2$ and magnetic induction ${\bf B}_2$ of the
charge $e_2$ are given by (4.2) and (4.3) respectively, but where the
quantities ${\bf\hat{n}}_2$, ${\bfvec{\beta}}_2$, $\dot{\bfvec{\beta}}_2$,
$R_2$ and $\kappa_2$ are referred to the retarded time $t_2$ of the charge
$e_2$ associated with the detection point ${\bf x}$ and time $t$.

In what follows we will carry out our calculations in the coordinate system
shown in Fig. 2; where the positions of charges $e_1$ and $e_2$ are described
by the vectors ${\bf r}_1(t)$ and ${\bf r}_2(t)$ defined in Eqs.  (3.2), and
the detection point is ${\bf x} = {\bf\hat{i}}\, r\sin \theta \cos
\varphi + {\bf\hat{j}} \, r \sin \theta \sin \varphi + {\bf\hat{ k}} \, r \cos
\theta$.  In Fig. 2 we have drawn the positions of the charges at three
differents times, namely at $t$, $t_1$ and $t_2$; where $t$ is the time at
which we are going to calculate the flux across the surface of the sphere of
radius $r > a$; the time $t_1$ corresponds to the retarded time of charge
$e_1$ associated with $({\bf x}, t)$, and $t_2$ is the retarded time of charge
$e_2$ associated with $({\bf x}, t)$.

\begin{center}
\fbox{\hspace{3cm}FIGURE 2 \hspace{6cm}}
\end{center}

Since the time interval $t - t_1$ that needs $e_1$ to go from its retarded
position $B_1$ to the actual position $A_1$ is the same that takes the light
for travel from $B_1$ to $P$, we have
\begin{equation}\label{e45}
t-t_1 = (a/c)\xi^{-1} \left\{ 1 + \xi^2 - 2 \xi \sin \theta \cos (\varphi -
wt_1)\right\}^{1/2}.
\end{equation}
Similarly, since the time interval $t - t_2$ that needs $e_2$ to go from its
retarded position $C_2$ to the actual position $A_2$, is the same that takes
the light for travel from $C_2$ to $P$, we have
\begin{equation}\label{e46}
t-t_2 = (a/c)\xi^{-1} \left\{ 1 + \xi^2 + 2 \xi \sin \theta \cos (\varphi -
\omega t_2) \right\}^{1/2},
\end{equation}
where $\xi$ denotes the parameter
\begin{equation}\label{e47}
\xi = \frac{a}{r} < 1.
\end{equation}
Equations (4.5) and (4.6) are complicated functional equations that determine
in an unique way the retarded times $t_1$ and $t_2$ respectively, as functions
of the parameters $t$, $r$, $\theta$ and $\varphi$.  The retarded times $t_1$
and $t_2$ are the same only for detection points over the $z$ axis.  Instead
of working with the retarded times $t_1$ and $t_2$, it is convenient to
introduce the following variables.
\begin{equation}\label{e48}
x = \varphi - \omega t_1,
\end{equation}
\begin{equation}\label{e49}
y = \varphi - \omega t_2.
\end{equation}

In the integral (4.1) the parameters $t$, $r$ and $\theta$ are fixed; then Eq.
(4.5) determines $t_1$ as a function of the angle $\varphi$, and therefore the
variable $x$ defined in Eq. (4.8) has a unique value for each $\varphi$ in the
interval $0 < \varphi < 2 \pi$.  This property allows us to carry out the
integral (4.1) as an integral over the variable $x$.  In order to see this
clearly, let us first note that the correspondence between the variables $x$
and $y$ in Eqs. (4.8), and (4.9) is one to one.  In fact, from Eqs. (4.5) and
(4.6) we get the following relation between $x$ and $y$
\begin{equation}\label{e410}
y - x = \beta\xi^{-1} \left\{ \left( 1 + \xi^2 + 2 \xi \sin \theta \cos
y\right)^{1/2} - \left( 1 + \xi^2 - 2 \xi \sin \theta \cos x\right)^{1/2}
\right\}
\end{equation}
where $\beta = a \omega/c$.  Taking the derivative with respect to $x$ in Eq.
(4.10), we obtain
\begin{equation}\label{e411}
\frac{dy}{dx} = \frac{1+ \rho_1^{-1}\beta \sin\theta\sin x}
{1+ \rho_2^{-1} \beta\sin\theta\sin y},
\end{equation}
where
\begin{equation}\label{e412}
\rho_1 = (1 + \xi^2 - 2\xi \sin\theta \cos x)^{1/2},
\end{equation}
and
\begin{equation}\label{e413}
\rho_2 = (1 + \xi^2 + 2\xi \sin \theta \cos y)^{1/2}.
\end{equation}
But
\begin{equation}\label{e414}
1 + \rho_1^{-1}\beta \sin \theta \sin x = \kappa_1 > 0,
\end{equation}
and
\begin{equation}\label{e415}
1 + \rho_2^{-1} \beta\sin \theta \sin y = \kappa_2 > 0,
\end{equation}
therefore
\begin{equation}\label{e416}
\frac{dy}{dx} = \frac{\kappa_1}{\kappa_2} > 0,
\end{equation}
which proves that the correspondence between $x$ and $y$ is one to one.  This
property holds for any time $t$, radius $r$ and angle $\theta$.

Let us consider now the integral (4.1); where the time $t$, the radius $r$ and
the angle $\theta$ remain fixed.  From Eq. (4.5) we obtain
\begin{equation}\label{e417}
-\omega \frac{dt_1}{d\varphi} =
\left(  \rho_1^{-1}\beta \sin \theta \sin x\right) \frac{dx}{d\varphi};
\end{equation}
if we combine this equation with Eq. (4.8), we get
\begin{equation}\label{e418}
\frac{dx}{d\varphi} = \frac{1}{\kappa_1} > 0,
\end{equation}
where $\kappa_1$ is explicitly given in Eq. (4.14).  Now, the equation
$d\varphi/dx = \kappa_1 > 0$ tells us that $\varphi$ is an strictly monotonous
increasing function of $x$; so we can put the integral (4.1) in the following
form
\begin{equation}\label{e419}
\int\limits_\alpha^{2\pi+\alpha} \, ({\bf E}_1 \times {\bf B}_2 + {\bf
E}_2 \times {\bf B}_1 ) \cdot {\bf{\hat{r}}} \kappa_1 dx,
\end{equation}
where the parameter $\alpha$ is given by
\begin{equation}\label{e420}
\alpha = - \omega t_1 (\varphi = 0) = - \omega t_1(\varphi = 2\pi),
\end{equation}
which in general depends in a complicated way on the time $t$, the radius $r$
and the angle $\theta$ of the band.  When the integrand of (4.19) is
explicitly evaluated by using the electric field (4.2) and the magnetic
induction (4.3), with the corresponding expression for $\bf{E}_2$ and
$\bf{B}_2$, it can be shown that the variables $x$ and $y$ appear only as
$\sin x$, \, $\cos x$, \, $\sin y$ and $\cos y$.  The integrand is, of course,
a function of the variable $x$ only since, as shown above, $y$ is uniquely
determined by the value of $x$ in Eq. (4.10).  Moreover, as it can be easily
seen, the correspondence between $x$ and $y$ is such that if $y$ is the value
associated with $x$, then $y + 2\pi$ is the value associated with $x + 2\pi$.
Thus we conclude that the integrand of (4.19) is a periodic function of $x$,
with a period of $2\pi$.  This property implies at once that the integral
(4.19) does not depend on the value of the parameter $\alpha$; so we can put
$\alpha = 0$ in it, obtaining the following representation for the energy-flux
(4.1).
\begin{equation}\label{e421}
\int\limits_0^{2\pi}\left( {\bf E}_1 \times {\bf  B}_2 + {\bf  E}_2 \times
{\bf
B}_1 \right) \cdot {\bf \hat{r}} \kappa_1 dx.
\end{equation}
In particular then, the energy flux across the band between $\theta$ and
$\theta + d\theta$ over the sphere of radius $r$ does not depend on the value
of the time at which it is evaluated.  The integral (4.21) depends, however,
in a very complicated way on the radius $r$ and the angle $\theta$.

The time-independence of the integral (4.21) is true for any band over the
surface of the sphere of radius $r$; therefore the energy flux across the
whole surface of the sphere is also independent of time. This property was
inferred on symmetry grounds in section II.  There we also proved that the
interference of radiation given by
\begin{equation}\label{e422}
(c/4\pi)r^2 \, \int\limits_0^\pi \,\sin\theta d\theta \,\int\limits_0^{2\pi}
\left( {\bf E}_1 \times {\bf B}_2 +
{\bf E}_2 \times {\bf B}_1 \right) \cdot {\bf{\hat{r}}} \kappa_1 dx
\end{equation}
can be evaluated for an arbitrary radius $r$ with $r > a$, because it is
independent of $r$.  The integrand of (4.22) contains a great number of terms
for any finite value of $r$, and in addition Eq. (4.10) that links the
variables $x$ and $y$ is very complicated for an arbitrary $r$.  Strong
simplifications of the integrand of (4.22) and of the functional relation
(4.10) are obtained when considering the limit when $r$ goes to infinity. We
emphasize the fact that, due to the independence of the interference rate on
the radius $r$, this limit does not present any complication, being perfectly
well defined.  In this limit Eq. (4.10) is reduced to
\begin{equation}
y - x = \beta \sin \theta (\cos y + \cos x),
\end{equation}
and the interference rate of radiation (4.22) becomes
\begin{equation}\label{e424}
-\frac{4}{3} \, \frac{e_1 e_2 c}{a^2} \, \beta^4 I(\beta),
\end{equation}
with $I(\beta)$ given by
\begin{equation}\label{e425}
I(\beta) = \frac{3}{4\pi} \, \int\limits_0^{\pi/2} \, \sin\theta d\theta
\, \int\limits_0^{2\pi} \,
\frac{(\cos^2\theta \cos x \cos y + \sin x \sin y - \beta^2 \sin^2 \theta)dx}
{(1 - \beta \sin \theta \sin x)^2 (1 + \beta \sin \theta \sin y)^3}.
\end{equation}

If instead of changing the variable $\varphi$ in the integral (4.1) by the $x$
of Eq. (4.8), we perform the integral (4.1) by means of the variable $y$
defined in Eq. (4.9), we would obtain in place of (4.25) the following
expression for $I(\beta)$.
\begin{equation}\label{e426}
I(\beta) = \frac{3}{4\pi} \, \int\limits_0^{\pi/2} \, \sin\theta d\theta
\, \int\limits_0^{2\pi} \,
\frac{(\cos^2\theta \cos x \cos y + \sin x \sin y - \beta^2 \sin^2 \theta)dy}
{(1 - \beta \sin \theta \sin x)^3 (1 + \beta \sin \theta \sin y)^2}.
\end{equation}
This formula looks different from (4.25), but it can be easily shown, with the
help of Eq. (4.23), that both formulae are the same.  We also point out that
the functional Eq. (4.23) has the solutions
\begin{equation}\label{e427}
x = y = \pi/2
\end{equation}
and
\begin{equation}\label{e428}
x = y = 3\pi/2
\end{equation}
independently of the value of $\beta$ and $\theta$.

Eq. (4.25), with $y(x)$ defined in Eq. (4.23), is an exact formula for the
interference radiation, and it is the main result of this paper.
Unfortunately, due to the complicated relation between the variables $x$ and
$y$ defined by Eq. (4.23), the integral (4.25) cannot be evaluated in a closed
analytical way.  This situation contrasts with what happens when we consider
the case of one charge in circular motion using the present treatment, where
the corresponding integral can be explicitly done as it is shown in appendix
B.

The integral (4.25) can be easily studied by means of numerical techniques. In
Fig. 3 we show the function $y(x)$ in the orbit plane, for three different
values of the parameter $\beta$; namely for $\beta = 0$,
\, $\beta = 0.5$ and $\beta = 0.9999$.  In Fig. 4 we present the results
of the numerical treatment of the function $I(\beta)$ of Eq. (4.25) in the
interval $0 \leq \beta < 0.99$ \cite{14}, where instead of the variable
$\beta$ we have used the variable $\phi$ defined by
\begin{equation}
\beta\cos \phi = \phi,
\end{equation}
since it is more convenient.  From the last equation we obtain
\begin{equation}
\frac{d\beta}{d\phi} = \frac{1+\beta\sin\phi}{\cos\phi}.
\end{equation}
The condition $\beta < 1$ implies $\phi < 0.739$, so that $d\beta/d\phi > 0$.
This shows that the correspondence between $\beta$ and $\phi$ is one to one.

In figure 4 we have also drawn a dotted curve that represents the power series
of Eq. (3.5)

\begin{center}
\fbox{\hspace{3cm}FIGURE 3 \hspace{6cm}}
\end{center}

\begin{center}
\fbox{\hspace{3cm}FIGURE 4 \hspace{6cm}}
\end{center}

\section{The interference rate in the ultrarelativistic case}
\setcounter{equation}{0}

When $\bfvec{\beta}$ is very close to one, the integrand of (4.25) is
significative only for values of the variables $x$ and $y$ around
\begin{equation}
x_1 = y_1 = \pi/2
\end{equation}

\begin{equation}
x_2 = y_2 = 3\pi/2
\end{equation}
In the approximate evaluation of (4.25) we will use a procedure similar to
that of reference \cite{15} for the one charge case; but now the
approximations are more crude because the integrand of (4.25) around (5.1) and
(5.2) is not as sharply defined as in the one electron case.  Nevertheless,
this somewhat heuristic procedure allows us to obtain a simple analytical
formula for the leading part of the interference rate, which accuracy improves
according as $\bfvec{\beta}$ becomes close to one.

Since the radiation is mainly concentrated in the orbit plane, it is
convenient to introduce the angle
\begin{equation}
\chi = \pi/2 - \theta
\end{equation}
Then, we are going to consider an expansion of the integrand of (4.25) with
$\delta x = x - x_1$ and $\chi$ of the order of $\gamma^{-1}$.   From (4.23)
it follows then that $\delta y = y - y_1$ is of the order of $\gamma^{-3}$,
that is, $y$ practically does not change when $x$ is around $\pi/2$, and we
can put $y = \pi/2$.  In this way we can approximate the integral of $x$
around $x_1$ in (4.25) by
\begin{equation}
\int\limits_{x_1 - \delta x}^{x_1  + \delta x}
\frac{[(1/2) (x - x_1)^2 - \alpha^2]dx}{2[(x - x_1)^2 + \alpha^2]^2}
\end{equation}
where
\begin{equation}
\alpha^2 = \gamma^{-2} ( 1 + [ \gamma \chi]^2).
\end{equation}
Extending the limits of integration between $-\infty$ and $+ \infty$ in (5.4),
we obtain for it the value $\pi/8 \alpha$.  Now, since the radiation is mainly
concentrated in the orbit plane, we can approximate $\sin\theta$ by $1$ in the
outermost integration of eq (4.25), and if the variable $x$ in changed by $z =
\gamma \chi$, we get the following contribution around $x_1 = \pi/2$,
for the interference rate of radiation.
\begin{equation}
I_1(\beta) = \frac{3}{32} \, \int\limits_0^{\gamma \pi/2} \, \frac{dz}{(1 +
z^2)^{1/2}} =
\frac{3}{32} \ln ( \gamma \pi)
\end{equation}
The contribution $I_2(\beta)$ around $x_2= 3\pi/2$ of the integral (4.25) must
be, on symmetry grounds, equal to (5.6).  In this way, we get the following
approximated formula for the interference rate of radiation in the
ultrarelativistic case.
\begin{equation}
-\frac{4}{3} \frac{e_1 e_2}{a^2}c \, \tilde{I}(\gamma)
\end{equation}
where
\begin{equation}
\tilde{I} (\gamma) = \frac{3}{16} \ln (\gamma \pi)
\end{equation}
In table 5.1 we represent the interference rate given by the approximated
formula (5.8), and the value of this quantity evaluated numerically from the
exact formula (4.25).  As expected, the accuracy of (5.8) improves when
$\gamma$ increases.

\[
\begin{tabular}{|r|c|c|r|}
\hline
\multicolumn{4}{|c|}{  Table 5.1}\\
\hline
\multicolumn{1}{|c|}{$\gamma$}&
\multicolumn{1}{|c|}{$I(\gamma)$}&
\multicolumn{1}{|c|}{$\tilde{I}(\gamma)$}&
\multicolumn{1}{|c|}{Accuracy}\\
\hline
100 & 1,235 & 1.0781 & -12,7 \% \\ 500 & 1,536 & 1,3799 & -10,16 \% \\ 1000 &
1,666 & 1,5098 & -9,37 \% \\ 5000 & 1,968 & 1,8116 & -7,95 \% \\ 10000 & 2,098
& 1,9416 & -7,45 \% \\ 30000 & 2,304 & 2,1476 & -6,79 \% \\
\hline
\end{tabular}
\]

The behavior (5.8) for the interference rate of radiation in the
ultrarelativistic case contrasts strongly with the behavior of the Larmor
terms of each charge, which behaves as $\gamma^4$ when $\gamma$ goes to
infinity.  In particular, in the ultrarelativistic case, the interference rate
of radiation is completely negligible in comparison with the Larmor terms of
each charge.

\section*{Acknowledgments}

We wish to thank I. Schmidt, and D. Walgraef, for useful discussion.  We also
want to thank FONDECYT whose support through project 92-0808 has been
determinant for the development of this research.

\newpage

\section*{APPENDIX A}

The far field produced by two charges in arbitrary motion can be expressed in
the following form: $$ {\bf E} = \sum\limits_{n=1}^\infty \, {\bf
E}_n\eqno{(A.1)} $$ where ${\bf E}_n$ is the last term in Eq. (3.1).
Introducing the notation
\addtocontents{equation}{A}
\renewcommand{\theequation}{A.\arabic{equation}}
\setcounter{equation}{1}

\begin{eqnarray}
{\bf r} &=& {\bf r}_1 = - {\bf r}_2 = a \cos \omega t \hat{i} + a \sin \omega
t \hat{j}\\ {\bf v} &=& {\bf v}_1 = -{\bf v}_2 = -a \omega \sin \omega t
\hat{i} + a
\omega \cos \omega t \hat{j}\\
\alpha &=& \hat{\bf{ r}} \cdot {\bf r} = a r \sin \theta \cos ( \varphi
- \omega t)\\
\eta &=& \hat{\bf{ r}} \cdot {\bf v} =
a \omega r \sin \theta \sin (\varphi - \omega t )\\
\hat{\bf {r}} & = & \sin\theta\cos\varphi \hat{i} + \sin\theta\sin\varphi\
hat{j} + \cos\theta\hat{k}
\end{eqnarray}
we obtain the following expressions for the fields:

\begin{eqnarray}
E_1 &=&
\frac{(e_1 - e_2)\omega^2}{rc^2} \left\{ {\bf r} - \alpha {\bf\hat r}\right\}
\\
E_2 &= &
\frac{2(e_1 + e_2)\omega^2}{rc^3} \left\{\eta {\bf r} + \alpha {\bf v} -
2 \alpha \eta {\bf\hat{r}} \right\} \\ E_3& = &
\frac{(e_1 - e_2)\omega^2}{2rc^4} \left\{(6\eta^2 - 7\omega^2 \alpha^2) {\bf
r} + 14 \alpha \eta {\bf v} +
(7 \omega^2 \alpha^3 - 20 \alpha\eta^2) {\bf\hat r} \right\}\\ E_4 &=&
\frac{4(e_1 + e_2)\omega^2}{3rc^5} \left\{(3\eta^3 - 12\omega^2 \alpha^
2\eta){\bf r} +
( 12 \eta^2 \alpha - 5 \omega^2 \alpha^3 ) {\bf v} \right. \nonumber\\ & & +
\left.(17 \omega^2 \alpha^3 \eta -
15 \alpha \eta^3){\bf\hat r}\right\}\\ E_5 &= &
\frac{(e_1 - e_2)\omega^2}{24rc^6} \left\{(120\eta^4 -
1080 \omega^2 \eta^2 \alpha^2 + 241 \omega^4 \alpha^4 ){\bf r} +
\right.\nonumber \\
& & \left.( 720 \eta^3 \alpha - 964 \omega^2 \alpha^3 \eta){\bf v} + \right.
\nonumber \\
& & \left.  (-840 \eta^4 \alpha + 2044 \omega^2
\eta^2 \alpha^3 - 241 \omega^4 \alpha^5){\bf\hat r} \right\}\\
E_6 &= &
\frac{(e_1 + e_2)\omega^2}{120 rc^7} \left\{(720\eta^5 -
12000 \omega^2 \alpha^2 \eta^3 + 8560 \omega^4 \alpha^4 \eta) {\bf r} +
\right.\nonumber \\
& & (6000 \alpha\eta^4 - 17120 \omega^2 \alpha^3 \eta^2 + 2256 \omega^4
\alpha^5 )
{\bf v} +\nonumber
\\
& & \left.( -6720 \alpha\eta^5 + 29120 \omega^2 \alpha^3
\eta^3 - 10816 \omega^4 \alpha^5 \eta) {\bf\hat r} \right\}
\end{eqnarray}
\newpage

\begin{eqnarray}
E_7 &= &
\frac{(e_1 - e_2)\omega^2}{720 rc^8} \left\{(5040\alpha^6 - 138600 \omega^2 \
alpha^2 \eta^4 +
209790 \omega^4 \alpha^4 \eta^2 - 19279 \omega^6 \alpha^6) {\bf r} +
\right. \nonumber\\
& & \nonumber \\ & & ( 55440 \alpha\eta^5 - 279720 \omega^2 \alpha^3 \eta^3 +
115674 \omega^4 \alpha^5 \eta) {\bf v} +\nonumber \\ & & \nonumber \\ & &
\left.(- 60480 \alpha\eta^6 + 418320 \omega^2 \alpha^3 \eta^4 - 325464 \
omega^4
\alpha^5 \eta^2 + 19279 \omega^6 \alpha^7 ) {\bf\hat r}
\right\}
\end{eqnarray}

\begin{eqnarray}
E_8 &= &
\frac{8(e_1 + e_2)\omega^2}{315 rc^9} \left\{(-10150\omega^6\alpha^6 \eta
+ 35280 \omega^4 \alpha^4 \eta^3 - 13230 \omega^2 \alpha^2 \eta^5 + 315
\eta^7 ) {\bf r} +\right. \nonumber \\
& & \nonumber \\ & & (-1957 \omega^6 \alpha^7 + 30450 \omega^4 \alpha^5 \eta^2
- 35280
\omega^2 \alpha^3 \eta^4 + 4410 \alpha \eta^6) {\bf v} +\nonumber  \\
& & \nonumber \\ & & \left.( 12107 \omega^6 \alpha^7 \eta - 65730 \omega^4
\alpha^5 \eta^3
+ 48510 \omega^2 \alpha^3 \eta^5 - 4725 \alpha \eta^7 ) {\bf\hat r} \right\}\\
E_9 &= &
\frac{(e_1 - e_2)\omega^2}{40320 rc^{10} } \left\{
(2771521 \omega^8 \alpha^8 - 55597920 \omega^6 \alpha^6 \eta^2 + 92786400
\omega^4 \alpha^4 \eta^4 - \right.\nonumber \\
& & \nonumber \\ && 22014720 \omega^2 \alpha^2 \eta^6 + 362880 \eta^8) {\bf r}
+ \nonumber \\ & & \nonumber \\ & & (-22172168 \omega^6 \alpha^7 \eta +
111195840 \omega^4 \alpha^5 \eta^3 - 74229120 \omega^2 \alpha^3 \eta^5 +
6289920 \alpha \eta^7 ) {\bf v}\nonumber \\ & & \nonumber \\ & & (-2771521
\omega^8 \alpha^9 + 77770088 \omega^6 \alpha^7 \eta^2 - 203982240 \omega^4
\alpha^5 \eta^4 +
96243840 \omega^2 \alpha^3 \eta^6 \nonumber \\ & & \nonumber \\ & & \left. -
6652800 \alpha \eta^8 ) {\bf\hat r} \right\} \end{eqnarray}

For circular motion the Pointying vector becomes $$
\bfvec{S} = \frac{c}{4\pi} E^2 {\bf\hat r}
\eqno{(A.15)}
$$ and it can be evalued using the expressions of this appendix.  If we now
perform an integration over the surface of a sphere centered in the origin and
that encloses the orbit, we obtain eq. (3.3).

\newpage

\section*{APPENDIX B}

In the case of one electron in circular orbit with constant velocity, the
energy flux across the spherical surface or radius $r$ and center at the orbit
center, namely $$ (cr^2/4\pi) \int\limits_0^\pi \sin\theta d\theta
\int\limits_0^{2\pi} (
{\bf E}_1 \times {\bf B}_1) \cdot {\bf\hat r} d\varphi,
\eqno{(B.1)}
$$ is, like Eq. (2.4), independent of the time $t$ and of the radius $r$ of
the sphere. On introducing in (B.1) the variable $x$ of Eq. (4.8), and on
taking the limit when $r$ goes to infinity, it becomes $$ (ce^2/4\pi a^2)
\beta^4 \int\limits_0^\pi \sin\theta d\theta
\int\limits_0^{2\pi} \left\{ \frac{1}{(1-\beta\sin\theta\sin x)^3} -
\frac{\gamma^{-2} \sin^2 \theta \cos^2 x}{(1-\beta\sin\theta\sin x)^5}
\right\} dx $$ These integrals, unlike the integrals of Eq. (4.25), can be
easily evaluated in a closed analytical way.  The result is, of course, Eq.
(3.4).

}

\newpage
\begin{center}
\section*{Figure and Table Captions}
\end{center}

{\bf Table 5.1. }\parbox[t]{12cm}{\setlength{\baselineskip}{1.5\baselineskip}
{Comparison between the numerical calculation of the exact expression (4.25),
here denoted by $I(\gamma)$, and the asymptotic formula (5.8), denoted by
$\tilde{I}(\gamma)$.  We also show the porcentual error with respect to the
numerical value.  As expected, the approximation (5.8) improves as $\gamma $
becomes larger.\\ }}

{\bf Figure 1.
}\parbox[t]{12cm}{\setlength{\baselineskip}{1.5\baselineskip}{Two charges
$e_1$ and $e_2$ in circular motion at constant angular velocity $\omega $.
The orbit has radius $a$, and the coordinate axes have been chosen so that the
orbit is centered at the origin and contained in the $X-Y$ plane.  We also
show two spherical surfaces $\Sigma_1$ and $\Sigma_2$ centered at the orbit's
center, $\Omega $ being the domain bounded by $\Sigma_1$ and $\Sigma_2$.\\ }}

{\bf Figure 2.
}\parbox[t]{12cm}{\setlength{\baselineskip}{1.5\baselineskip}{At the
observation, time $t$ the charges $e_1$ and $e_2$ are ubicated at opposite
ends of diameter $A_1 A_2$.  In the same way, at the retarded times $t_1$ and
$t_2$ of charges $e_1$ and $e_2$ they are ubicated at the ends of the dotted
diameter $B_1 B_2$ and the dashed diameter $C_1 C_2$ respectively.  We also
show the retarded distances $R_1$ and $R_2$ from the retarded positions of the
charges to the observation point $P$, and the radius $r$ of the spherical
surface to which $P$ belongs.  \\ }}

{\bf Figure 3.
}\parbox[t]{12cm}{\setlength{\baselineskip}{1.5\baselineskip}{The curves
represent the function $y(x)$ defined by the retardation condition, eq.
(4.23).  The plot is made for three different values of the parameter $\beta$,
namely \, $\beta = 0$, \, $\beta= 0.5$ and $\beta = 0.9999$.  Note that in all
cases $x = y = \pi/2$ and $x = y = 3\pi/2$ are solutions of eq. (4.23).  \\ }}

{\bf Figure 4. }\parbox[t]{12cm}{\setlength{\baselineskip}{1.5\baselineskip}{
The solid lines show the numerical evaluation of function $I(\beta)$ defined
by eq. (4.25).  The plot is made in terms of the variable $\phi$ defined by
$\beta\cos\phi = \phi$, for the range $0 <
\beta \leq 0.99$.  We also show a dotted line that represents the power
series expansion for the interference rate, eq. (3.5), in the range $0 <
\beta \leq 0.4$.           }}

\end{document}